# Enhanced Figure of Merit in Bismuth-Antimony Fine-Grained Alloys at Cryogenic Temperatures

Sheng Gao[1, a)], John Gaskins[2], Xixiao Hu[1], Kathleen Tomko[2], Patrick Hopkins[2, b)], S. Joseph Poon[1, c)]

[1] Department of Physics, University of Virginia, Charlottesville, VA 22904-4714
[2] Department of Mechanical and Aerospace Engineering, University of Virginia, Charlottesville, VA 22904-4259

**Abstract:**

Thermoelectric (TE) materials research plays a vital role in heat-to-electrical energy conversion and refrigeration applications. Bismuth-antimony (Bi-Sb) alloy is a promising material for thermoelectric cooling. Herein, a high figure of merit, ZT, near 0.6 at cryogenic temperatures (100-150K) has been achieved in melt-spun n-type $Bi_{85}Sb_{15}$ bulk samples consisting of micron-size grains. The achieved ZT is nearly 50% higher than polycrystalline averaged single crystal ZT of ~0.4, and it is also significantly higher than ZT of less than ~0.3 measured below 150K in Bi-Te alloys commonly used for cryogenic cooling applications. The improved thermoelectric properties can be attributed to the fine-grained microstructure achieved from rapid solidification, which not only significantly reduced the thermal conductivity but also mitigated a segregation effect. A record low thermal conductivity of ~1.5 W m$^{-1}$ K$^{-1}$ near 100 K was measured using the hot disk method. The thermoelectric properties for this intriguing semimetal-semiconductor alloy system were analyzed within a two-band effective mass model. The study revealed a gradual narrowing of the band gap at increasing temperature in Bi-Sb alloy for the first time. Magneto-thermoelectric effects of this Bi-Sb alloy further improved the TE properties, leading to ZT of about 0.7. The magneto-TE effect was further demonstrated in a combined NdFeB/BiSb/NdFeB system. The compactness of the BiSb-magnet system with high ZT enables the utilization of magneto-TE effect in thermoelectric cooling applications.

a) sg5jk@virginia.edu

b) peh4v@virginia.edu

c) sjp9x@virginia.edu

**Introduction:**

Renewable energy has become an important topic in modern society for addressing sustainable global energy demand. Thermoelectric (TE) technology enables the conversion of currently under-utilized waste thermal energy directly into electrical power as well as having the potential for cooling and refrigeration applications[1–4]. In order to increase the energy conversion efficiency of TE devices, a good dimensionless figure of merit, ZT, of TE materials is required which is defined by the following equation,

$$ZT = \frac{\sigma S^2}{\kappa}T, \qquad (1)$$

where $\sigma$ is the electrical conductivity, $S$ is the Seebeck coefficient, $\kappa$ is the total thermal conductivity consisting of electrical, phononic, and bipolar parts at temperature $T$. Due to the physical correlation between these parameters, it is usually difficult to obtain a high Seebeck coefficient and electrical conductivity simultaneously while keeping a low thermal conductivity.

Among many different thermoelectric materials, bismuth antimony (Bi-Sb) alloys, known as the first topological insulator[5], stood out as a promising cryogenic temperature TE material. Both pure Bi and Sb elements are semi-metals, with inverted $L_a$ and $L_s$ band positions[6]. Upon adding Sb to Bi to form a solid solution, band inversion starts and opens a gap between Sb concentrations of 7%-22% with a maximum value of ~ 30meV at 15-17% Sb[7]. Being a narrow-gap semiconductor, at low temperature, Bi-Sb has the potential to outperform other ordinary TE materials such as $Bi_2Te_3$, which is one of the best TE materials near room temperature[8,9].

There have been several studies of the thermoelectric properties of single crystal Bi-Sb since the 1960s[10–16]. In particular, Yim and Amith[14] investigated the temperature dependent properties of different single crystal Bi-Sb compositions along two crystalline axes (trigonal and binary/bisectrix axes in a rhombohedral $A_7$ structure of the space group $R\bar{3}m$, as shown in Fig. 1(a)). In its undoped state, Bi-Sb is an n-type semiconductor. Yim and Amith also investigated the magneto-thermoelectric effects in Bi-Sb, which gave rise to enhancement of the Seebeck coefficient in magnetic fields. The effects were explained on the basis of the transverse-transverse thermo-galvanomagnetic effects which are the results of co-action between the Hall effect, Nernst effect and Righi-Leduc effect. They reported ZT~0.55 along the trigonal axis of Bi-Sb in the absence of a magnetic field around 80K but ZT peaked at a value of ~0.4 along the binary/bisectrix axis at 150K. By applying an external magnetic field of 3 kOe, ZT could increase to ~1 along the trigonal axis. On the other hand, magneto-TE effects along the binary/bisectrix directions were not reported. After more than four decades, this high ZT remains the highest among commonly utilized cryogenic TE materials, such as those based on $Bi_2Te_3$ with ZT below 0.3 at 100-150K[17,18]. Furthermore, since the ZT of $Bi_2Te_3$ alloys peaks near 300-400K, while it is near 100-150K for BiSb, Bi-Sb and Bi-Te can be combined to form the segmented element of a TE cooler module to enhance the coefficient of performance of a TE cryocooler.

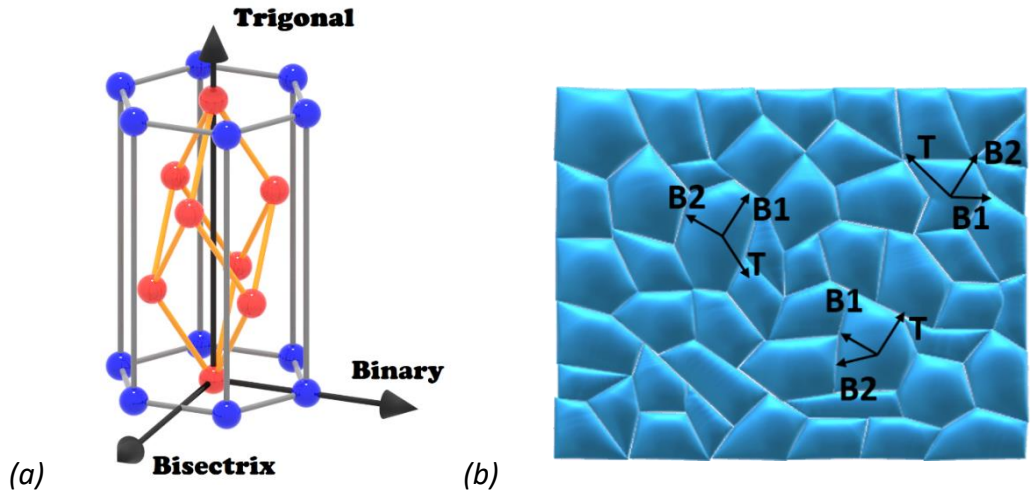

*Fig.1 (a) Rhombohedral crystal structure in a hexagonal representation. (b) Schematic representation of polycrystalline Bi-Sb material where the blocks represent the poly-grains with different crystal orientations of T, B1, and B2, which stand for Trigonal, Bisectrix,and Binary, respectively.*

There are technical hurdles to overcome before Bi-Sb alloys can be utilized as a practical material for cooling applications. A well-known drawback of single crystals is poor mechanical property. In order to pursue more robust alloys, efforts have also been made to manufacture polycrystalline Bi-Sb[19–23]. A simplified representation of a polycrystalline Bi-Sb alloy is shown in Fig. 1 (b). Due to the large temperature gap between the solidus and liquidus lines in the Bi-Sb phase diagram, significant phase segregation always occurs under normal cooling conditions, preventing the formation of a homogeneous alloy ingot upon cooling down from the liquid mixture phase. To overcome this issue, mechanical alloying has been preferentially used by researchers in order to improve the compositional homogeneity of the samples[21,22]. However, longtime high energy ball milling is prone to contamination, which can have adverse effects on TE properties. Other attempts to accelerate the cooling process, such as liquid-nitrogen quenching, have also been reported. However, the quenched ingots still require a time-consuming furnace anneal to homogenize after fabrication[19]. One group reported using melt-spinning for a higher cooling rate, which yielded variable properties as a function of linear rotation speed, indicating that the TE properties might not be optimized[20].

In the present study, $Bi_{85}Sb_{15}$ was selected for its optimal band structure to maximize the power factor between 100K and 150K. Through the melt-spinning (MS) method with a rapid cooling rate near $10^6$ K/s, followed by spark plasma sintering (SPS), a homogeneous nanocrystalline Bi-Sb alloy was prepared. The melt-spun samples result in a low thermal conductivity measured using the hot disk method[24]. Thermal transport measurements carried out via hot disk effectively avoid the heat loss issues experienced with Physical Property Measurement Systems (PPMS) due to large metal contacts. Additionally, hot disk is insensitive to heat capacity as it is a direct measurement of thermal conductivity[24,25]. In this study, a high ZT~0.55 was obtained near 125K in fine-grained Bi-Sb in the absence of magnetic fields.

**Experiments method:**

Randomly mixed Bismuth and Antimony pieces with purity of 99.999% were sealed into a vacuum quartz tube and heated up to 800 °C in a furnace for 2 hours. The liquid Bi-Sb solution was quenched into liquid nitrogen and the as-quenched ingots were made into thin ribbons by a melt-spinning (MS) method with different linear wheel speeds ranging from 15m/s - 40m/s. Our measurements show TE properties are almost independent of wheel speed, which means 15m/s already gives a sufficiently high cooling rate. Ribbons were ground into fine powders in a mortar followed by consolidation via a spark plasma sintering (SPS) system (Thermal Technology LLC 10-4 model) at 240°C under 50MPa of pressure for 15 minutes. Investigation of microstructure and compositional homogeneity were performed by a PANalyticalX'Pert pro-MPD X-ray diffraction (XRD) instrument and a FEI Quanta 650 in both scanning electron microscope (SEM) and energy-dispersive X-Ray spectroscopy (EDS) configurations. Samples were cut into cuboid bars with dimensions of 15mm×2.5mm×1.5mm and measured by a PPMS from Quantum Design to obtain resistivity and Seebeck coefficients using the four leads method. The hot disk method was used for thermal conductivity measurements. The Hall coefficient was measured using the Van der Pauw method on a VersaLab system from Quantum Design. The N42 grade NdFeB magnet plates used in the combined BiSb/magnet system were obtained from K&J Magnetics.

**Results and discussion:**
**(1) Structural characterizations**

Fig. 2 shows the XRD pattern of a $Bi_{85}Sb_{15}$ sample after undergoing SPS. Single phase Bi-Sb solid solution with a rhombohedral $A_7$ structure of space group $R\bar{3}m$ was confirmed, indicating a solid solution was well formed.

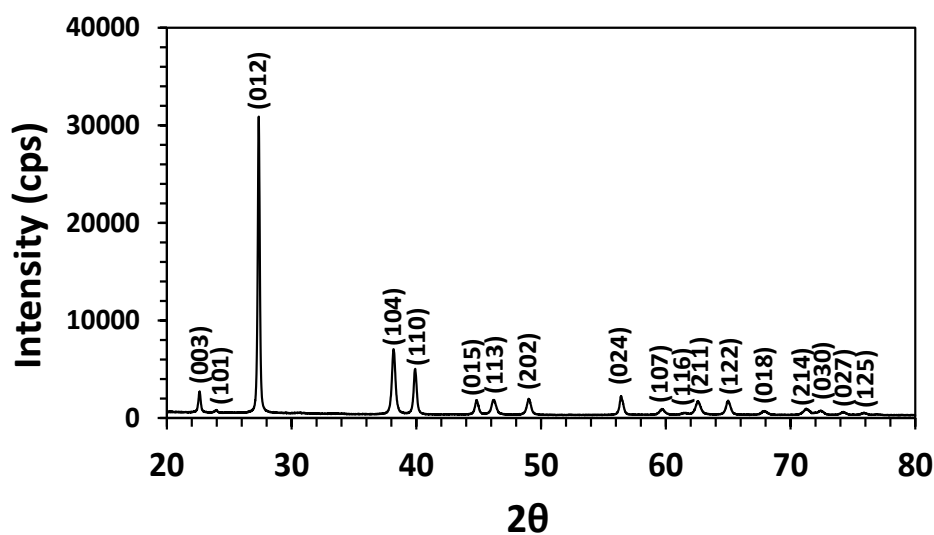

*Fig. 2   X-ray diffraction pattern of a $Bi_{85}Sb_{15}$ sample after SPS. The diffraction peaks are indexed using the ICDD XRD profile for Bi since Sb and Bi are isostructural elements.*

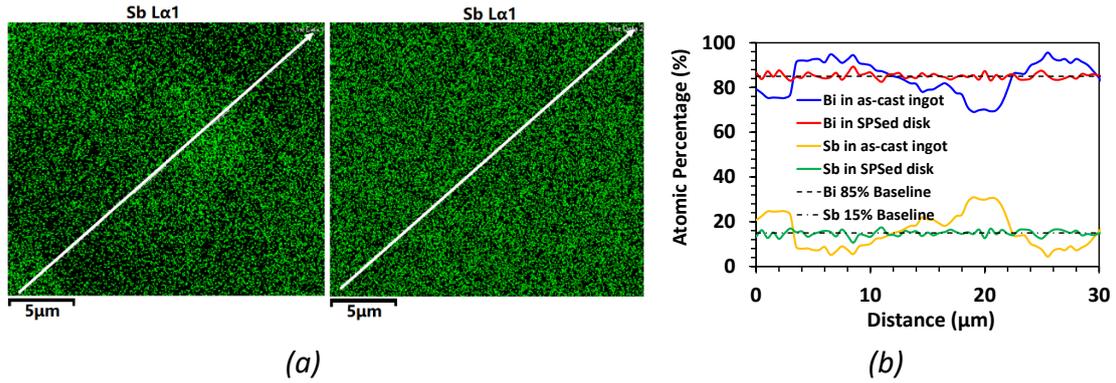

*(a)*             *(b)*

*Fig. 3 (a) Sb concentration mapping of $Bi_{85}Sb_{15}$ using energy-dispersive spectroscopy (EDS). Left: As-cast ingot; Right: SPS sample. (b) Elemental concentration of Sb obtained by scanning along the white lines shown in (a). Bi concentration profile was obtained using the corresponding Bi concentration map.*

Energy-dispersive X-Ray Spectroscopy (EDS) was used to examine sample homogeneity. Sb concentration mappings were performed on an as cast ingot and an SPS sample. In Fig. 3 (a), the density of green dots in an area represents the concentration of Sb content. In the as-quenched ingot, where a clear inhomogeneous distribution can be seen by eye, which was confirmed via a line scan at intervals of 25 microns as shown in Fig. 3(b). This inhomogeneity is present even after a liquid nitrogen quench, which means appreciable phase segregation can occur on the order of seconds, causing large concentration differences up to nearly 30% at different spots. In contrast, the Sb concentration distribution for the sample after melt-spinning and spark plasma sintering showed a much smaller range of Sb concentration variation, under 1 at.% Sb, as shown in Fig. 3 (b), indicating an effective homogenization was achieved through rapid cooling by melt-spinning.

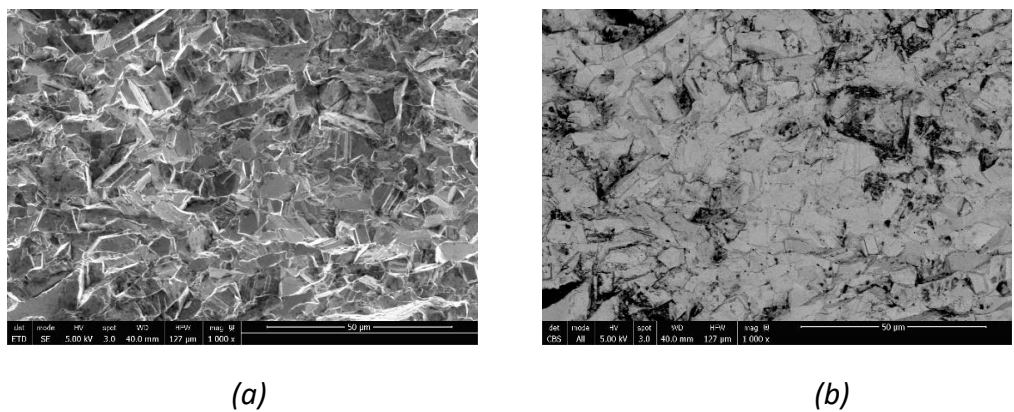

*(a)*             *(b)*

*Fig. 4 Micro-structure images taken by scanning electron microscope (SEM) on the cross section of SPS $Bi_{85}Sb_{15}$ sample. The scale bar is 50 µm. (a) image taken from the Everhart-Thornley detector (ETD); (b) image taken from the concentric ring backscatter detector (CBS).*

The surface of Bi-Sb sample was well polished to provide a smooth finish in preparation for micro-structure study using scanning electron microscopy. Fig. 4 shows the SEM images for the SPS $Bi_{85}Sb_{15}$ sample, fine grains of micron size can be seen. The ETD image shows the surface information of the cross section while the CBS image shows the topographical details. Electron backscatter diffraction (EBSD) images could be found in the supplementary materials for further grain information. By fracturing the sample, we also found nano-size particles that seem to have been embedded among the micron-size matrix grains. The microstructure observed by us is similar to that reported by Luo et al[20].

**(2) Thermoelectric transport properties**

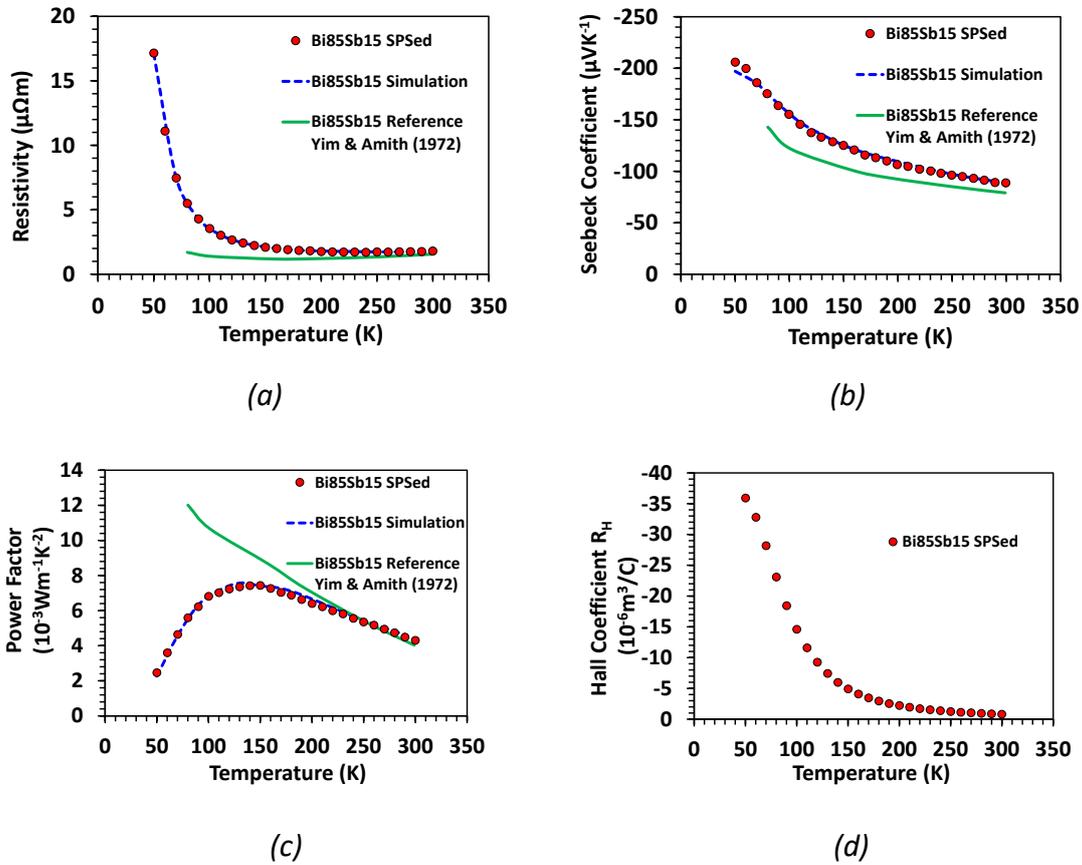

*Fig. 5   Resistivity (a), Seebeck coefficient (b), power factor (c), and Hall coefficient (d) results for the SPS $Bi_{85}Sb_{15}$ sample. For parts (a), (b), and (c), the red circles denote experimental data measured with PPMS, blue dashed lines represent simulation results from the two-band effective mass model, and green solid lines are the literature values from the averaged single crystal reference experiments by Yim and Amith.*

Fig. 5 shows resistivity, Seebeck coefficient, power factor, and Hall coefficient as measured by PPMS. The results for single crystals reported by Yim and Amith are shown for comparison as solid green lines[14]. In order to directly compare their results with our polycrystalline sample, an average along three orthogonal axes (trigonal, binary, and bisectrix axes) was taken of the single crystal results for electrical conductivity, Seebeck coefficient, and thermal conductivity[14]. At low temperature, the electron-dominated transport infers an electron carrier density of ~$10^{17}$ cm$^{-3}$ and high mobility of ~$10^4$ cm$^2$/V.s. As discussed below, the large electron carrier mobility plays an important role in the magneto-thermoelectric effect in Bi-Sb. The fast increase in resistivity of the SPS Bi$_{85}$Sb$_{15}$ fine-grained sample is likely caused by carriers scattering at grain boundaries as the mean free path of carriers increases to the micron size at low temperatures in this high mobility solid solution alloy. The latter leads to an apparent increase in the bandgap relative to that observed in single crystals, but comparable to those reported for other fine-grained samples[7]. The higher resistivity in Bi-Sb also resulted in an increase in the Seebeck coefficient as seen in other thermoelectric alloys[26]. The negative Hall coefficient trends towards zero at increasing temperature, indicating the increasing role of hole carriers.

**(3) Two-band effective mass model analysis**

To quantitatively understand the thermoelectric transport properties, resistivity and Seebeck coefficient were simulated using a two-band effective mass model based on the Boltzmann transport equation[27].

A parabolic band assumption was applied, and the L$_a$ band was the conduction band while the H band was the valence band. The density of states can be written as:

$$D(E) = \frac{1}{2\pi^2}\left(\frac{2m_d^*}{\hbar^2}\right)^{\frac{3}{2}}\sqrt{E} \tag{2}$$

where $m_d^*$ is the total density of states effective mass and $E$ is the electron energy.

From the Boltzmann transport equations, we get the expressions for electrical conductivity and the Seebeck coefficient:

$$\sigma = -\frac{2e^2}{3m_c^*}\int_0^\infty \frac{\partial f}{\partial E}D(E)E\tau dE \tag{3}$$

$$S = \frac{1}{eT}\frac{\int_0^\infty \frac{\partial f}{\partial E}D(E)E(E-E_f)\tau dE}{\int_0^\infty \frac{\partial f}{\partial E}D(E)E\tau dE} \tag{4}$$

where $e$ is the unit charge, $m_c^*$ is the conductivity effective mass, $E_f$ is the chemical potential, $\tau$ is the scattering time and $f$ is the Fermi-Dirac distribution function.

Within the two-channel transport model, we can write the resistivity $\rho$ and Seebeck coefficient $S_{total}$ as:

$$\frac{1}{\rho} = \sigma_{total} = \sigma_n + \sigma_p \tag{5}$$

$$S_{total} = \frac{\sigma_n S_n + \sigma_p S_p}{\sigma_n + \sigma_p} \tag{6}$$

where the subscripts n and p above refer to the electron and hole channel components of the electrical conductivity and Seebeck coefficient, respectively.

The distance between the chemical potential and the bands will be affected if the band gap changes. The subsequent variation in the band gap may have effects on TE transport properties. Details where $\sigma_{n,p}$ and $S_{n,p}$ are the electrical conductivity and Seebeck coefficient components of the electron channel and hole channel, respectively.

The distance between the chemical potential and the bands will be affected if the band gap changes. The subsequent variation in the band gap may have effects on TE transport properties. Details of the calculation and simulation parameters can be found in the Supplementary Materials. In order to fully simulate the TE transport properties, it is necessary to allow a temperature dependent band gap. Results are shown in Fig. 6. Band gap narrowing at increasing temperature is conceivable in the case of Bi-Sb alloys with a small band gap and robust semimetal-to-semiconductor crossovers. Furthermore, a lattice deformation that was observed near 150 K could also modify the band gap[28].

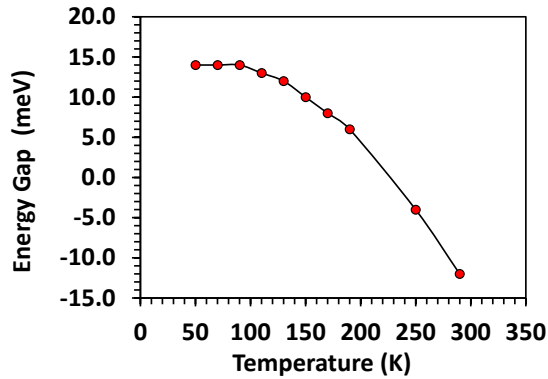

*Fig. 6  Energy gap obtained from the two-band effective mass model analysis of thermoelectric properties Bi$_{85}$Sb$_{15}$. Solid line is guide to the eye.*

**(4) Magneto-thermoelectric effect and its realization in portable magnetic field**

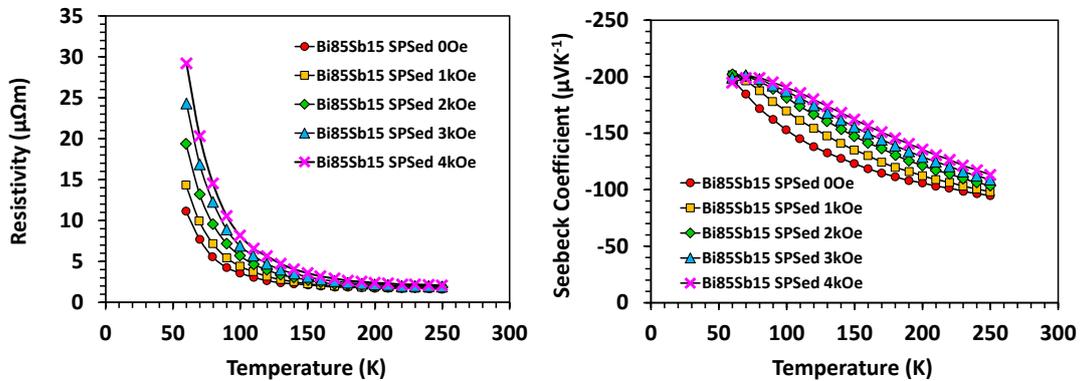

*Fig. 7 Magneto-resistance and magneto-thermopower of SPS Bi$_{85}$Sb$_{15}$ sample measured under different magnetic fields.*

Bi-Sb single crystals have exhibited high magneto-thermoelectric effects, whereby the thermopower measured along the trigonal axis increased when a magnetic field (B) was applied along the bisectrix direction[11,14]. In Fig. 7, magneto-resistance and magneto-TE properties in magnetic fields up to 4 kOe were shown for our $Bi_{85}Sb_{15}$ sample. A consistent increase in $\rho$ with B was observed for all temperature ranges. At the same time, enhancement in the Seebeck coefficient was also found above 50K. This magneto-TE effect could be explained on the basis of the transverse-transverse thermo-galvanomagnetic effects,[11,14] which are the result of interactions between the Hall, Nernst, and Righi-Leduc effects.

For a simplified discussion, considering orthogonal heat flow and uniform magnetic field in an isotropic polycrystalline Bi-Sb alloy, the thermopower enhancement has a quadratic dependence on the applied magnetic field given by :

$$\Delta S = B^2(NR_H\sigma + L_R N) \qquad (7)$$

where N is the Nernst coefficient, $R_H$ is the Hall coefficient, $L_R$ is the Righi-Leduc coefficient, $\sigma$ is the electrical conductivity and $\Delta S$ is the change of measured adiabatic Seebeck coefficient in a magnetic field[14] . As noted by Yim and Amith, for the Bi-Sb system, $L_R \ll R_H\sigma$, therefore only the first term in equation (7) needs to be considered. In the weak field limit, the Nernst coefficient can be written as N~$\mu/\varepsilon_F$, where $\mu$ is the carrier mobility and $\varepsilon_F$ is the Fermi energy[29,30]. $\Delta S$ in equation (7) can therefore be approximated as $\Delta S$~ -$\mu\sigma B^2$. Thus, it can be seen that the very high electron mobility has led to the high Nernst coefficient in Bi-Sb alloys, bringing about the appreciable increase in thermopower observed. Meanwhile, the magneto-resistance which is given approximately by the relation $\Delta\rho$~$\mu B^2$, [29] also increases due to the high carrier mobility. In lieu of measurement of thermal conductivity in a magnetic field, we can only estimate the optimal ZT, which occurs around 3 kOe based on our thermoepower and electrical resisitivity measurement.

In order to exploit the observed magneto-TE effect in a versatile thermoelectric cooling device, the device must necessarily be physically compact. For the present Bi-Sb alloys, it is possible to generate a moderate magnetic field near 3 kOe in the TE sample with a pair of NdFeB magnet plates. As shown in Fig. 8, two NdFeB magnet plates of 0.8mm thickness with magnetization near 1.3 T were configured parallel to each other with the 2-mm wide Bi-Sb sample sandwiched in between. The simulated magnetic field profile inside the sample showed a rather uniform magnetic field distribution in the range 3 – 3.3 kOe, except for the small areas near the edges. The details of the magnetic field simulation can be found in the Supplementary Materials.

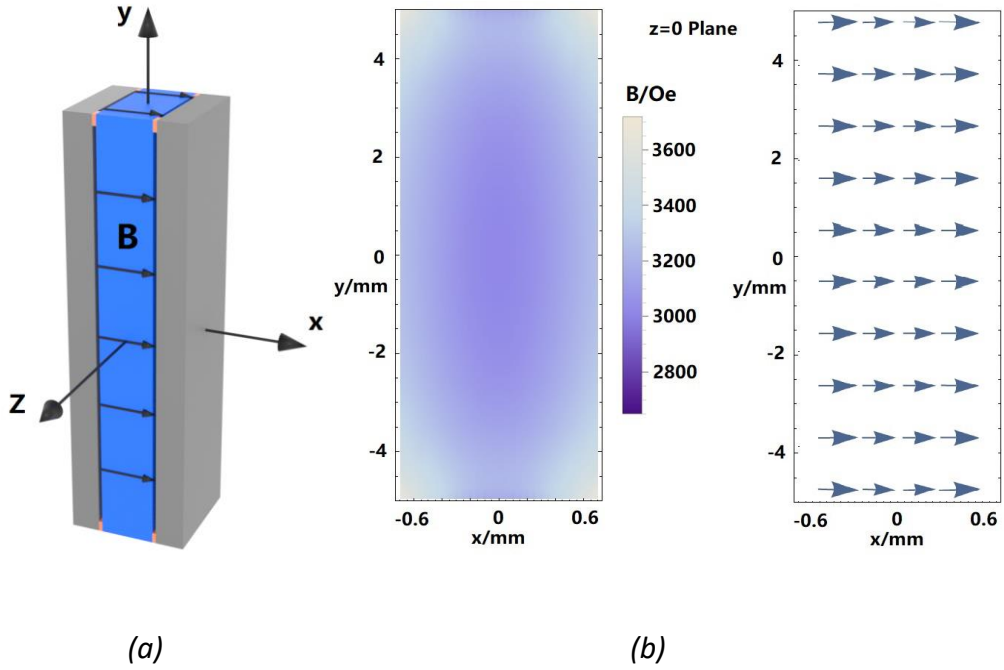

*(a)*                            *(b)*

*Fig. 8 (a) BiSb sample (blue) and NdFeB magnet plates (grey) setup and (b) simulated magnetic field profile. The Origin O is located at the center of the sample. The strength and direction distributions of the magnetic field in the z=0 plane are shown in the center and right illustrations, respectively.*

In our thermoelectric measurement using PPMS, part of the surface area of the sample was occupied by thermal and electrical contacts made of copper rings. Therefore, only the middle half of the sample was covered by a pair of magnetic plates, as shown in Fig. 9 (a). The magnetic plates were fixed to the sample with four narrow strips of double-sided insulating tape. Thus, a thin gap was left in between each magnetic plate and the sample to minimize the heat loss in the Seebeck coefficient measurement. The probe configuration used in the PPMS measurement is shown in Figure 9 (b). Measurement in the absence of an applied magnetic field was performed without the magnet plates. The TE properties were measured for the sample segment between the two thermometers attached to copper rings, $C_1$ and $C_2$. The sample segment can be divided into three sections connecting in series: (1) Middle section in magnetic field, with length $L_M$, (2) Upper separated section in near-zero field, with length $L_B$, and (3) Lower section identical to the upper separated section. Therefore, the thermoelectric properties of the middle section between the pair of magnetic plates can be calculated based on the measurement. The details of the calculations can be found in the Supplementary Materials.

The results of the estimated thermoelectric properties between the 3kOe magnet plates are shown in Fig. 10. It shows good agreement with the data in uniform 3 kOe fields obtained using PPMS.

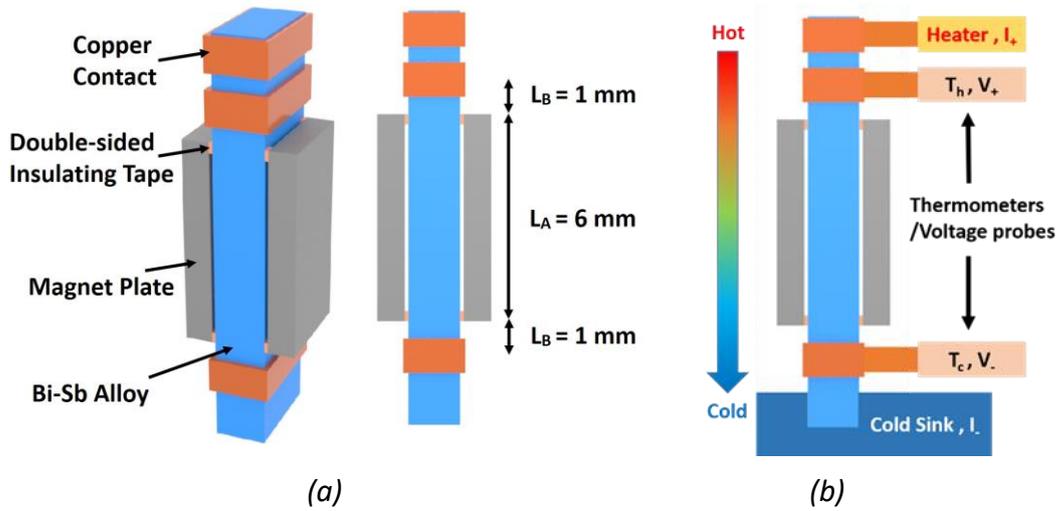

*(a)* *(b)*

*Fig. 9 (a) Schematic of sample and magnetic plates arrangement employed to demonstrate that magneto-thermoelectric effect can also be realized in a compact combined BiSb/magnet system. (b) Schematic of the probe configuration for thermal transport measurement using the PPMS.*

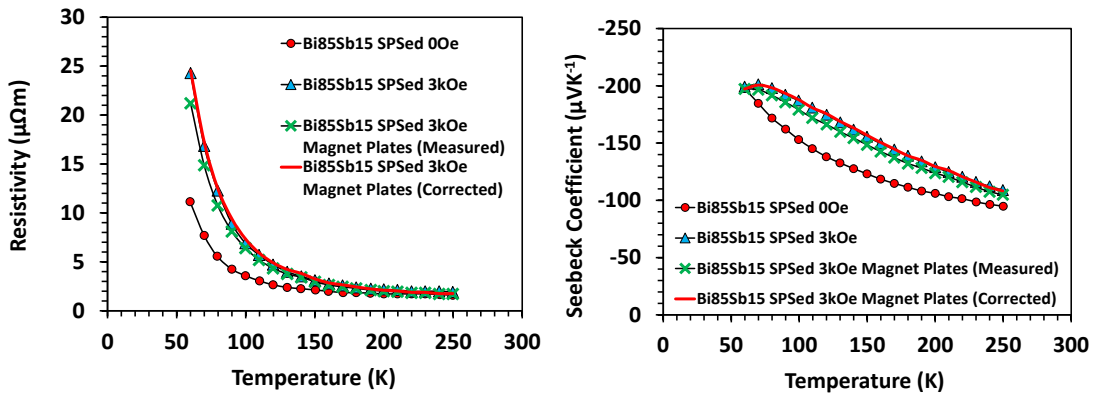

*Fig. 10 Magneto-resistivity and magneto-Seebeck coefficient of SPS $Bi_{85}Sb_{15}$ measurements without applying an external magnetic field and in a magnetic field of 3 kOe produced by the PPMS superconducting magnet or NdFeB magnetic plates. Red circle, blue triangle, and green cross each represents measurement without applying a magnetic field, in 3 kOe generated by the PPMS magnet, and in magnetic field produced by the NdFeB plates, respectively. Red solid line represents magneto-TE results calculated for the middle section of the sample shown in Fig. 9 (c/o Supplementary Materials).*

**(5) Thermal conductivity and figure of merit ZT**

Among TE properties, thermal conductivity is commonly the most difficult to measure accurately. Typical methods used in previous studies relied mostly on the heat flow across the whole sample with multiple high thermal conductivity metal contact leads attached. Unsystematic variations of thermal conductivity data have been observed in measurements using PPMS, and previous studies by other groups relying on this type of measurement also reported a wide range of thermal conductivities[20,21,31,32]. This indicates there could be significant uncertainty associated with the thermal conductivity measurements due to radiative heat loss during the heating event or effects from metal contacts, especially for a low thermal conductivity system such as Bi-Sb. As a result, the transient plane source technique, often referred to as the hot disk method, was used to obtain temperature dependent measurements of thermal conductivity on bulk samples without the aforementioned test concerns[24,25].

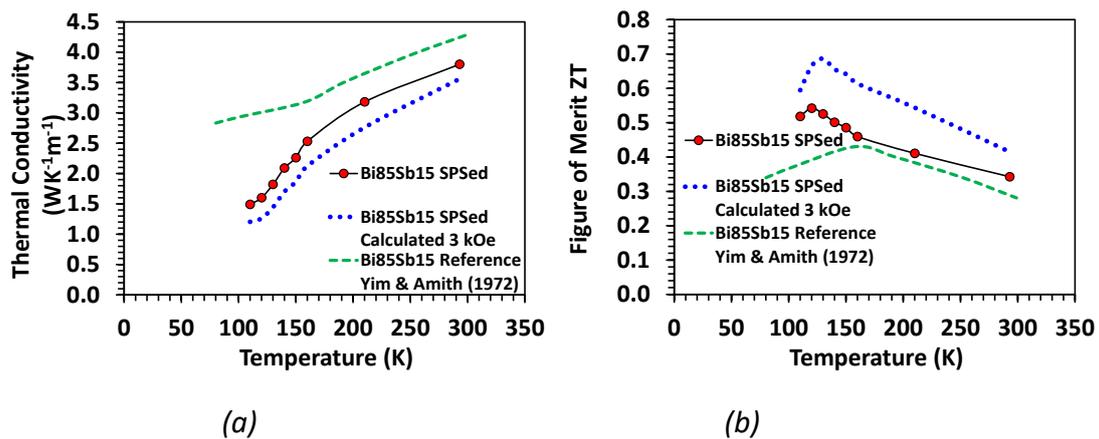

(a) (b)

*Fig. 11  (a) Red circle solid line represent thermal conductivity of SPS $Bi_{85}Sb_{15}$ without an applied magnetic field measured using the hot disk method. Blue symbols represent the thermal conductivity calculated using Wiedemann–Franz law assuming that the sample is measured in a magnetic field of 3 kOe. Green dashed line represent the reference data for the polycrystalline averaged single crystal measurements on $Bi_{85}Sb_{15}$ reported by Yim and Amith. (b) The dimensionless figure of merit, ZT, are obtained using the thermal conductivity results shown in (a) and the electrical resistivity and thermopower data from Fig. 5 and Fig. 10.*

The thermal conductivity of samples was measured via hot disk from 110K to room temperature using a cold stage heat sunk to liquid nitrogen.   Hot disk uses an electrically conductive pattern encapsulated by an insulating material, in this case Kapton, to both source and sense an electrically generated heat source.   Placing the sensor between two identical samples allows us to assume the heat source provided by the sensor is located in a semi-infinite medium.   The solution to the heat equation in this case yields an absolute measurement of the thermal conductivity of the sample.   This is a key advantage over other thermal conductivity methods which often rely on measurements of thermal diffusivity or effusivity and calculate thermal conductivity by having *a priori* knowledge of the heat capacity of the material.   This aspect of the measurement technique becomes increasingly important as we measure samples which do not have vetted literature values for heat capacity at room temperature or low temperatures such as those investigated in this study.

Temperature-dependent thermal conductivity data of SPS $Bi_{85}Sb_{15}$ without applied magnetic field measured using hot disk are presented in Fig. 11 together with the calculated dimensionless figure of merit, ZT, using the power factor from Fig. 5. The results are compared with polycrystalline averaged reference data[14]. The thermal conductivity of ~1.5 W m$^{-1}$ K$^{-1}$ near 100 K is the lowest reported for an undoped Bi-Sb system[14–16,20–23].

The total thermal conductivity consists of three parts:

$$\kappa_{total} = \kappa_{electronic} + \kappa_{lattice} + \kappa_{bipolar}. \qquad (8)$$

The electronic contribution to thermal conductivity, $\kappa_{electronic}$, is frequently described using Wiedemann-Franz law:

$$\kappa_{electronic} = L\sigma T \qquad (9)$$

where L is the Lorenz number.

The Lorenz numbers used in the calculation were obtained from the Seebeck coefficient based on the empirical equation proposed by H. S. Kim etc.[33] ($L\sim1.8\times 10^{-8} W\Omega K^{-2}\ W$ for our sample and $L\sim1.85\times 10^{-8} W\Omega K^{-2}\ W$ for Yim & Amith's sample). This Lorenz number model shows good agreement with the values obtained from band structure calculations with scattering assumptions. Even for the systems with multiple non-parabolic bands, it can be regarded as a good approximation. Typical measurements of $\kappa_{total}$ and ZT are acknowledged to have uncertainties near 10 and 15 %, respectively[34].

At low temperatures (T<150K), where bipolar terms are relatively small and can be ignored in Eq. 8 due to the lack of hole carriers, $\kappa_{lattice}$ can be estimated by subtracting $\kappa_{electrical}$ from $\kappa_{total}$. It is found that our SPS Bi$_{85}$Sb$_{15}$ has a $\kappa_{lattice}\sim0.85\ W K^{-1}m^{-1}$ near 100K, which is significantly lower than the value $\kappa_{lattice}\sim1.5\ W K^{-1}m^{-1}$ calculated by the same method for the polycrystalline averaged single crystal results of Yim and Amith[14]. Thus, the large reduction of $\kappa_{lattice}$ has played a key role in suppressing the $\kappa_{total}$. This reduction is speculated to be associated with the fine-grained microstructure of the Bi-Sb samples. The presence of nano-size particles in unpolished fractured sample could also facilitate multi-scale phonon scattering,[35] which deserves more detailed study in the future.

Even though the current thermal conductivity measurement does not offer the capacity to directly measure thermal conductivity in a magnetic field, it is noted that the magnetic field dependence of κ is essentially determined by the magnetic field dependence of κ$_{electronic}$ due to the large magneto-resistivity. As such, the Wiedemann–Franz law can be used again to calculate κ$_{electronic}$, which is combined with the $\kappa_{lattice}\sim0.85\ W K^{-1}m^{-1}$ obtained above to calculate κ$_{total}$. As shown in Fig. 11(a), the κ$_{total}$ of SPS Bi$_{85}$Sb$_{15}$ in 3 kOe field is seen to decrease by 15-25 percent at 100-150K compared with that measured without applying a magnetic field. It is worth pointing out that we have also measured the thermal conductivity of our Bi-Sb samples in 3 kOe using the PPMS. A similar percentage decrease in κ$_{total}$ was observed. This lends some support to our calculation even though PPMS does not give the correct magnitude of thermal conductivity. κ$_{total}$ is combined with the electrical resistivity and thermopower in 3 kOe from Fig. 10 to estimate the figure of merit ZT in 3 kOe field, which is presented in Fig. 11 (b).

Conclusion:

In summary, the thermoelectric properties of n-type fine-grained Bi-Sb alloys have been investigated. The alloys showed a figure of merit ZT near 0.6 at ~125K in zero magnetic field and up to 0.7 in a moderate in-situ magnetic field produced with a compact combined NdFeB/BiSb/NdFeB system. The obtained ZT values significantly exceeded those of polycrystalline averaged BiSb single crystal value of ~0.4 as well as Bi2Te3 alloys with ZT~0.3 most often used for TE cooling applications. A low thermal conductivity was obtained by fabrication via rapid solidification, which created a fine-grained structure. A two-band effective mass model was employed to simulate the thermoelectric transport properties, which uncovered a heretofore unreported narrowing of the indirect band gap with increased temperature. The versatility of the compact TE-magnetic system makes it possible to couple n-type BiSb with a p-type cryogenic thermoelectric material in a cooling device.

**Acknowledgments:**

This work was supported by the Defense Advanced Research Projects Agency MATRIX Program contract HR0011-16-C-0011 (P.I. Dr. Rama Venkatasubramanian, Johns Hopkins University Applied Physics Laboratory). The content of the information does not necessarily reflect the position or the policy of the Government, and no official endorsement should be inferred. Approved for public release; distribution is unlimited. The authors also thank Dr. Normand Modine (Sandia National Laboratories) for discussion.

**Data availability statement:**

The data generated and/or analyzed during the current study are available from the corresponding author on reasonable request.

**Author contributions statement:**

S. G. performed electrical transport and thermopower measurements and wrote the first draft of the manuscript, both S. G. and X. H. participated in the materials synthesis and carried out the two-band model simulation; J. G. performed the thermal conductivity measurements and edited the manuscript; Kathleen Tomko performed the scanning electron microscopy study, P. H. supervised the thermal conductivity measurements; S. J. P. supervised the project and edited the manuscript. All authors participated in the discussion.

**Competing Interests**

The authors declare no competing interests.